\documentclass[12pt,halfline,a4paper]{ouparticle}
\usepackage{dcolumn}
\usepackage{bm}
\usepackage{mathptmx}
\usepackage{feynmf}
\usepackage{tikz-feynman}
\usepackage{amsmath}
\usepackage{amssymb}
\usepackage{mathtools}
\usepackage{calrsfs}%%%%%%%%%%<------add
\DeclareMathAlphabet{\pazocal}{OMS}{zplm}{m}{n}%%%%%%%%%%<------add
\usepackage{caption}
\usepackage{color}
\usepackage{listings}
\usepackage{dirtytalk}
\usepackage{url}
\usepackage{cite}
\usepackage{mathrsfs}
\usepackage{threeparttable}
\usepackage{adjustbox}
\usepackage{cancel}
\usepackage[english]{babel}
\usepackage{hyperref}
\hypersetup{
    colorlinks=true,
    linkcolor=red,
    citecolor=blue,
    filecolor=red,      
    urlcolor=red,
    }
\renewcommand{\thesection}{\Roman{section}} 
\renewcommand{\thesubsection}{\Roman{subsection}}
\begin{document}
\title{Neutrino mass and lepton-asymmetry in \break {Scotogenic model}}

\author{%
\name{Sumit Satapathy}
\emph{\address{University of Delhi, Department of physics and astrophysics, New-Delhi,\\
110021, India}
}}
\abstract{We start out with the extension of the Standard Model with fermions $N_i(i=1,2,3)$, which are odd under $Z_2$ symmetry and singlet under $SU(2)_L$, with the SM-like fermions we also have a doublet scalar field $\Phi_2$ like Higgs but odd under discrete symmetry. This model has potential similar to the inert doublet model, which implies we can have a Dark Matter candidate similar to IDM \cite{gandma}, the Dark matter aspect and calculation of DM parameters are not discussed extensively. In this model, we can calculate the mass of neutrinos from loop corrections \cite{zee1980theory} that were massless at the tree level, and define a physical mass matrix which requires the inclusion of the Weinberg operator and we get a mass term as, $m_D^TM^-m_D$.  Then we look at lepton number violation through decays of heavy Majorana neutrinos, these generate a lepton asymmetry, then with the help of sphalerons they convert the excess of lepton number to baryon number, this conversion is quantified by $a_{sph}$. To calculate the number densities we use the Boltzmann equation, and $CP$ asymmetry parameter $\epsilon$ which gives us the net lepton number produced, at last, the present baryon-to-photon ratio is estimated.}
\date{}

\maketitle

\subsection{\label{intro}Introduction}
Over the past few decades, the work on the extension of the Standard Model of Particle physics has been of priority for physicists. This mainly includes Dark matter which accounts for five times the abundance of Baryonic matter in our universe, various evidence for DM \cite{jungman} lies in the observation of rotation curves, cluster formations, Cosmic microwave background, and the data from the Planck satellite. Adding to the observations for Dark matter, Neutrino oscillation was observed from neutrinos coming directly from the sun (solar neutrinos) which led the way for the possibility of masses for neutrinos that are different from other members of the lepton family.\\
An additional puzzle that is bothering physicists is the asymmetry between matter and anti-matter known as Baryogenesis. To explain baryogenesis there are certain conditions that are needed to be fulfilled, known as Sakharov conditions \cite{sakharov1998violation} which have been elaborated on in the coming section. Baryogenesis explains why there is an abundance of baryons over its counterpart, although they started with similar equilibrium abundances as predicted, to explain this we rely on out-of-equilibrium decays. One such process is Leptogenesis \cite{buchmuller2004some},  this process is used to explain the observed asymmetry in baryon abundance through the help of lepton number-violating processes. Leptogenesis creates a lepton number asymmetry through out-of-equilibrium decays at a particular scale/temperature that is converted to baryon asymmetry through specific electro-weak transitions, known as sphalerons \cite{kuzmin1985anomalous}. These transitions conserve $B-L$ as a whole and violate them individually, making these global symmetries ($B$ , $L$) anomalous. Also, in addition to explaining Dark matter and asymmetry in the universe, This model explains the generation of neutrino masses \cite{ma2006common}, we know that to explain the generation of neutrino masses we use the see-saw mechanism \cite{boucenna2014low,amsler2015nuclear}. The most simple and elegant way to get a nonzero eigenvalue for mass is the LR symmetric model where we add a right-handed fermion for their left-handed partners, this gives a mass matrix with a nonzero diagonal term that is Majorana mass term, depending on the values for  $n \times m $ Higgs field we have different types of seesaw mechanism. In this model we explain neutrino masses from radiative corrections, in which we generate a similar term for fermionic mass from one loop correction to self-energy diagram of left-handed neutrinos, this process includes integrating out the right-handed neutrino from our Lagrangian \cite{ma1998pathways}, we talk about it in more detail in coming sections. Where this framework is able to explain these phenomena, the potential part of the Lagrangian with scalar fields can also assign a DM candidate to the lightest neutral component of the second Higgs doublet.  \\
In the course of this literature, we concentrate on the aspect of the generation of neutrino masses and lepton number asymmetry in the scotogenic framework. We start by defining the field content for the building of the scotogenic model in sec (\ref{model}), sec (\ref{mass}) is devoted to how we generate neutrino mass from radiative corrections with the use of loop and we show how with the help of a higher dimension operator we can get our desired. In sec (\ref{lepto}) we discuss the method of leptogenesis through which the observed baryon to photon ratio can be obtained, we conclude in sec (\ref{concl}) about the discussions done in the entirety of this literature.

\subsection{\label{model}Scotogenic model}
The minimal scotogenic model includes additional fermionic fields $N_i$ (\textbf{1},\textbf{1},\textbf{0}) \cite{ma2001naturally} where in the minimal case we take $i=\text{3}$ ($i=$generation), in addition to the fermionic field we also have a Higgs doublet $\Phi_2$ (\textbf{1},\textbf{2},\textbf{1/2}) they differ from the SM fermion in the sense that they are odd under a discrete symmetry $Z_2$ which can be understood as a subgroup of continuous gauge symmetry. This symmetry can have two possible values odd and even, the standard model is even under this $SM_{field}\xrightarrow{DS}SM_{field}$. And the additional fields are odd under this transformation as $N_i\xrightarrow{DS}-N_i$, due to this additional symmetry the lightest neutral component of the second Higgs doublet is a suitable candidate for Dark matter, as it cannot decay further.
Let us see the potential sector with the scalar fields \cite{Satapathy:2023qts}, 
\begin{equation*}
V=m_1^2{|\Phi_1|^2}+m_2^2{|\Phi_2|^2}+\lambda_1{|\Phi_1|^4}+\lambda_2{|\Phi_2|^4}+\lambda_3{|\Phi_1|^2|\Phi_2|^2}+\lambda_4{(\Phi_1^+\Phi_2)(\Phi_2^+\Phi_1)}
\end{equation*}
\begin{equation}
\label{1}
    +\frac{\lambda_5}{2}{((\Phi_1^+\Phi_2)^2+ h.c)}
\end{equation}
Due to $Z_2$ symmetry, we do not see any trilinear couplings, also the 2HD has a zero-vev after the EWSB, and using unitary gauge condition we get,\\\\
\centerline{$\Phi_1$=$
\begin{pmatrix}
    0\\
    \frac{1}{\sqrt{2}}(v+h)\\
\end{pmatrix}
$          \space\space\space                    $\Phi_2$=$\begin{pmatrix}
    {H}^+\\
    \frac{1}{\sqrt{2}}(H+iA)\\
\end{pmatrix}
$}
\\\\where $h$ is SM Higgs boson, $H$ and $A$ are neutral scalar field and $H^+$ is charged field. The physical masses can be calculated from the potential after imposing symmetry breakdown and they are as follow,
$$m^2_h=2\lambda_1{} v^2$$
$$m^2_{H^{\pm}}=\frac{1}{2}\left(\lambda_{3}v^2-m_2^2\right)$$
$$m^2_A=m^2_{H^{\pm}}+\frac{1}{2}\left(\lambda_4-\lambda_5\right)v^2$$
$$m^2_H=m^2_{H^{\pm}}+\frac{1}{2}\left(\lambda_4+\lambda_5\right)v^2$$
We take the lightest neutral component and our DM candidate as $H$, the small non-degeneracy between mass for $H$ and $A$ is also kept intact. So from (\ref{eq2}) we can put additional constraints on the coupling as $\lambda_5<0$ and $\lambda_{45}<0$. The total number of free parameters is five in this scalar potential.
\begin{equation}
\label{eq2}
    m_H<m_A,m_{H^{\pm}}
\end{equation}
 The parameters related to Dark matter such as its relic density from the Boltzmann equation and direct detection through scattering cross-sections can be obtained, but we do not elaborate on them in this literature, but can be found in \cite{jungman,1,15,2,9}.
 In the next section, we go through the process for the mass generation of neutrinos in the scotogenic model through radiative corrections. It is best to explain beforehand that, the simplest mechanism for neutrino mass generation (type-1 seesaw) lacks behind in the aspect of testability when compared to the radiative models \cite{merle2015running}, which generate mass at the loop level.
\subsection{\label{mass}Neutrino mass at one-loop level}
The Lagrangian with the $Z_2$ odd fermions is \cite{borah2019tev},
\begin{equation}
\label{eq1}
    \mathcal{L}_{f}=\frac{1}{2}\left(\overline{N}_{i}M_{ij}N^{C}_j+H.c\right)+\left(Y_{ij}\Bar{L}_i\Tilde{\Phi}_2N_j + H.c\right)
\end{equation}
The Yukawa term includes the left-handed fermions from the standard model, $Z_2$-odd singlet fermions, and added scalar doublet. To obtain the mass for an electron in SM we have the term as, $\mathcal{L}_{yukawa}=-y\Bar{L}\Phi_1e_R$, after symmetry breaking we get the mass. With this construction, the charged leptons and down-type quarks get masses. For the remaining fermions such as neutrinos(or the up-type quark), we use,
$$\Bar{L}\Phi_1\longrightarrow\Bar{L}\sigma_2\Phi_1^*$$
where $\Tilde{\Phi}_n=i\sigma_2\Phi_n^*$, its also easy to see that the above term is $SU(2)$ invariant, therefore you will find this sort of term in (\ref{eq1}). In SM we do not have a discrete symmetry therefore if there exists a right-handed neutrino we can have a Dirac mass term but due to the unbroken $Z_2$ symmetry, the active neutrinos cannot have a Dirac mass term from $\Bar{L}\Tilde{\Phi}_1N$ and hence remains massless at tree level.
It is interesting to know that we do not need right-handed neutrinos to give neutrinos mass. If we allow the non-renormalizable term of dim-5, which can be generated by integrating out the right-handed neutrino from the Lagrangian \cite{putnam2019neutrino}, this term is called the Weinberg operator \cite{weinberg1979baryon}. 
$$\mathcal{L}_{dim-5}=-\overline{M}_{ij}\left(\Bar{L}^i\Tilde{\Phi}\right)\left(\Tilde{\Phi}\Bar{L}^j\right)^\dagger$$
If the mass-eigenstate sterile right-handed neutrino is very heavy, a dim-3 mass term is indistinguishable from the dim-5 mass term. We can therefore generate neutrino masses without the addition of right-handed neutrinos.
There are methods through which we can generate our small neutrino mass from the dim-5 operator, the operator after EWSB takes a similar form of mass term for fermions \cite{cai2018lepton},
  \begin{equation*}
   \mathcal{L}^{(5)}=\frac{C_5}{\Lambda}(L\Phi)(L\Phi)
   \xrightarrow{\text{EWSB}} C_5\frac{v_0^2}{2\Lambda}\cdot\overline{(v_L)^c}v_L
  \end{equation*}
large $\Lambda$ with $C_5 \approx 1$ (seesaw models) \cite{cai2017trees},
$$\frac{v^2}{\Lambda} \approx 0.05\text{eV}\implies\Lambda\approx 10^{14}\text{GeV}$$
Small $C_5$ with $\Lambda>v$ (inverse seesaw, radiative models$\dots$),
$$C_5\frac{v^2}{\Lambda}\approx \frac{v^2}{\Lambda}\left( \frac{1}{16\pi^2}\right)^n\prod_{i=1}y_i$$
For small $C_5$ and low $\Lambda$, these models are more testable than the first case with $\Lambda$ in the GUT scale. We should remember that the above operator and approximations are for Majorana masses.
 \\
 In SM after the EWSB the scalar field has a vev associated with it and in unitary gauge, the operator has a generic form of $M\Bar{\psi}\psi$. But due to our discrete symmetry not being broken, the second Higgs doublet has zero-valued vev, and after EWSB, at tree level we do not get a mass term as earlier, from this operator. So at tree level, as there is no Dirac mass linking, the neutrinos remain massless at tree level. We can remember from one loop correction, mass renormalization in QED, that we get a mass correction in terms of bare and renormalized parameters, we can do a similar operation for the neutrino self-energy diagram. From (\ref{eq1}) and (\ref{1}) we can find out the possible interaction vertices (3-point and 4-point) scalar vertices. Therefore the one-loop correction to neutrino self-energy is,
$$
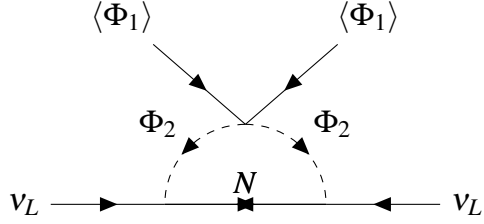
\begin{tikzpicture}
\begin{feynman}
         \vertex(a);
    \vertex[ left=of a](b){$\nu_L$};
    \vertex[above right =of a](d);
        \vertex[ below right=of d](c);
  \vertex[  right=of c](e){$\nu_L$};
  \vertex[above left= of d](f){$\langle \Phi_1\rangle$};
  \vertex[above right= of d](g){$\langle \Phi_1\rangle$};
    
    \diagram{
    (b) -- [fermion](a);
    (e) -- [fermion](c);
    (d) -- [dashed, fermion , quarter right,edge label'=$\Phi_2$](a);
    (d) -- [dashed, fermion , quarter left, edge label=$\Phi_2$](c);
    (a) -- [fermion,edge label=$N$](c);
    (c) -- [fermion,edge label'=$N$](a);
    (f) -- [fermion](d);
    (g) -- [fermion](d);

     };
     \end{feynman}
     \end{tikzpicture}$$
The below diagram has physical mass eigenstates \cite{escribano2020generalizing}, now the 1-loop integral takes the form, also the term with $\cancel{k}$ doesn't contribute because it is odd in the momentum integral, so the 1-loop integral becomes more like a one-loop correction to the two
point function of general $\phi^4$ theory, a similar operation can be done for two loop diagram \cite{babu1988model}. \\
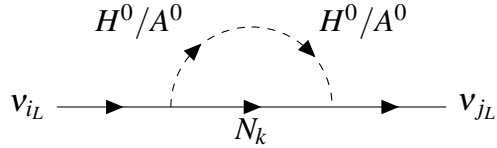
 $$ \begin{tikzpicture}
\begin{feynman}
         \vertex(a);
    \vertex[ left=of a](b){$\nu_{i_L}$};
    \vertex[ right=of c](d){$\nu_{j_L}$};
    \vertex[above right= of a](e);
     \vertex[below right =of e](c);
    \diagram{
    (b) -- [fermion](a);
    (a) -- [dashed, fermion, quarter left,edge label=$H^0/A^0$](e);
    (e) -- [dashed, fermion, quarter left,edge label=$H^0/A^0$](c);
    (c) -- [fermion](d);
    (a) -- [fermion,edge label'=$N_k$](c);
   
      };
     \end{feynman}
     \end{tikzpicture}$$
$$I\sim\int\frac{\text{d}^4k}{2\pi^4}\frac{i}{k^2-m^2_{H^0/A^0}}\left(\frac{i\left(\cancel{k} +M_{k}\right)}{k^2-M^2_{k}}\xrightarrow{}\frac{iM_k}{k^2-M^2_k}\right) $$\

 $$\left(M_\nu\right)_{ij}= \sum\frac{Y_{ik}Y_{jk}M_k}{16\pi^2}\left [\frac{m^2_{H^0}}{m^2_{H^0}-M^2_k}\ln\frac{m^2_{H^0}}{M^2_k}- \frac{m^2_{A^0}}{m^2_{A^0}-M_k^2}\ln\frac{m^2_{A^0}}{M_k^2}\right]
$$\\
It is cumbersome to evaluate the convergent integrals, and the calculations have various approximations to take a simpler form.
The left-handed neutrinos interact via the weak force, so it is natural to include flavor basis than mass basis, in flavor basis the couplings to boson are diagonal but not mass, then the mass eigenstates are related to flavor by PMNS matrix \cite{avila2022revisiting}, these are related by unitary transformations.\\
$m^2_{H^0}-m^2_{A^0}=2\lambda_5v^2$ is clearly small when compared to $m^2_{H^0}+m^2_{A^0}=2m^2_\mu$ due to constraints on $\lambda_5$ which gives \cite{ma2006verifiable},
$$\left(M_\nu\right)_{ij}=\frac{\lambda_5v^2}{8\pi^2}\sum\frac{Y_{ik}Y_{jk}M_k}{m^2_\mu-M^2_k}\left[1-\frac{M^2_k}{m^2_\mu-M^2_k}\ln\frac{m^2_\mu}{M_k^2}\right]$$
Now if $m^2_\mu\simeq M_k^2$ then,
$$\left(M_\nu\right)_{ij}\simeq \frac{\lambda_5v^2}{16\pi^2}\sum\frac{Y_{ik}Y_{jk}}{M_k}$$
When we integrate out the right-handed neutrino and put it in (\ref{eq1}) we get the mass as \cite{casas2001oscillating}, where $Y$ is the Yukawa matrix,
\begin{subequations}
\begin{equation}
    M_\nu=Y^T_\nu M^{-1} Y_\nu(v)^2
\end{equation}
\begin{equation}
    \zeta=\frac{M_\nu}{v^2}=Y^T_\nu M^{-1} Y_\nu
\end{equation}
As mentioned earlier we can always diagonalize our mass matrix with the unitary PMNS matrix$\left(v_{Lj}=U^{ij}v_{Lj}\right)$ as,
\begin{equation}
    U^T\zeta U=\text{diag}(\zeta_1,\zeta_2,\zeta_3)=D_\zeta
\end{equation}
After Using the Casas-Ibarra parametrization, to produce the physical masses, the Yukawa matrix can be written as,
\begin{equation}
    Y_\nu=D_{\sqrt{M}}OD_{\sqrt{\zeta}}U^+
\end{equation}
 where $O$ is an orthogonal matrix,
\begin{equation}
    O=D_{\sqrt{M^{-1}}} Y_\nu UD_{\sqrt{\zeta^{-1}}}
\end{equation}

\end{subequations}
\subsection{\label{lepto}Leptogenesis}
The abundance of matter over antimatter or baryon asymmetry can  be explained through a process called leptogenesis. It is important to know about a few other things before we explain leptogenesis. The SM has global symmetry as in Baryon and lepton number, which comes from the invariance in phase rotation. But these symmetries are anomalous in nature $\partial_\mu\langle J^{\mu n} \rangle\not=0$, $B$ and $L$ numbers are individually violated , but $B-L$ is not as $(\partial_\mu J^B_\mu=\partial_\mu J^L_\mu)$ \cite{buchmuller2005leptogenesis}. Due to this, certain configurations preserve $B-L$ but violate $L$ and $B$ \cite{harvey1990cosmological}, such configurations are known as sphalerons \cite{kuzmin1985anomalous}. Now coming back to lepton asymmetry, the heavy Majorana RH neutrinos can decay as, $N\xrightarrow{}l\Phi_2$ and $N\xrightarrow{}\Bar{l}\Phi_2^*$ these create SM leptons, these are out of equilibrium decays. The lepton asymmetry is converted to baryon asymmetry through sphalerons transitions, which explains the baryon asymmetry, known as Baryogenesis, GUT scale analysis of baryogenesis is given in \cite{kolb2018early}. Now to understand sphalerons let me give you a familiar analogy, when we have a double potential in single-dimension space time, the solution to the time-independent finite energy E-L equation when wick rotated mimics the tunneling and in that classical field theory the inverted potential that we get from making our path integral convergent gives me euclidean instantons. Now to go from one vacuum to another there should be tunneling, but the probability of tunneling is very small. But in a thermal bath where fluctuations can overcome the barrier and these transitions can happen, this is the KRS mechanism \cite{kuzmin1985anomalous,klinkhamer1984saddle}.
Certain conditions needed to be satisfied for the generation of matter-antimatter asymmetry known as Sakharov conditions, and our model satisfies those conditions. It should be noted that as we will be talking about higher temperatures such that the only mass possible is for the heavy neutrinos and zero mass for LH fermions.\\
So the process for the generation of asymmetry is as follows \cite{fukugita1986barygenesis}, at a scale $T>M_i$ there exists $\Delta B \not = \Delta L \not = 0$ the excess of lepton number is from GUT scale, the excess of lepton number and baryon is washed out by equilibrium process and electroweak phase transitions. Now when the process falls out of equilibrium (when the expansion rate is greater than $\Gamma$ or $T<M_i$) we have newly created $\Delta L\not = 0 $. From the sphalerons process the excess of the lepton number is converted into the excess of the baryon number \cite{davidson2002lower} keeping $B-L$ constant. And as the universe cools down and the Higgs field comes into action and we get the observed asymmetry. \\ In this sec (\ref{lepto}) we define a parameter $\epsilon$ which explains the net lepton number produced in the decays and with a non-zero value explains the excess of the lepton number. It should be remembered that we follow the mass hierarchy of $M_1<M_2,M_3$, and $N_1$ being the lightest odd particle cannot decay which is similar to DM candidate, therefore we consider decays of $N_2$ for generated asymmetry \cite{mahanta2019fermion,antusch2012fuller}. After that we use the Boltzmann equation to calculate a particular species evolution, then we solve these equations for solutions for number density in comoving coordinates. At last, we relate the excess of the lepton number to the baryon number through sphalerons. This is the overview of this section. Now let us start with the Boltzmann equation\footnote{The notation for number density follow \cite{kashiwase2012baryon} and a general idea about Boltzmann equation can be seen in \cite{kolb2018early}, we use these equations for both Dark matter abundance as well as for baryogenesis} that governs the time evolution of Heavy Majorana neutrinos \cite{buchmuller2005pedastrian},
\begin{equation}
\label{3}
    \frac{dY_N}{dx}=-D\left(Y_{N}-Y^{eq}_{N}\right)
\end{equation}
\begin{equation}
\label{4}
    \frac{dY_{B-L}}{dx}=-\epsilon_ND\left(Y_{N}-Y^{eq}_{N}\right)-W^{Total}{Y_{B-L}}
\end{equation}
$$D=\frac{\Gamma_N}{Hx}=K_{N}x\frac{K_1(x)}{K_2(x)}$$
Where $K_i(x)$ is the modified Bessel function of $i^{th}$ kind,
$$K_i(x)=\frac{\sqrt{\pi}}{\Gamma\left(i+\frac{1}{2}\right)}\left(\frac{x}{2}\right)^i\int_{1}^{\infty}e^{-zx}\left(z^2-1\right)^{i-\frac{1}{2}}\text{d}z$$
and decay parameter as,
$$K_N=\frac{\Gamma_2}{H(x=1)}$$
Where $N_Z(t)=n_Z(t)R_*(t)^3$ is the particle number in some comoving volume which contains one photon at time $t_*$, the amount of $B-L$ asymmetry as $Y_{B-L}$ and $CP$ asymmetry parameter as $\epsilon_N$.
Eq.(\ref{3}) tells about the evolution of our heavy neutrinos due to decay and inverse decays $N\leftrightarrow l\Phi_2$ and $N\leftrightarrow\Bar{l}\Phi^*_2$, and Eq.(\ref{4}) has all the process that violates lepton number and $B-L$ because after dropping of temperature this number density freezes in to be the observed asymmetry, These includes the decay and inverse decays of $N_i$ and $\Delta L=2$ scattering terms, We are not elaborating on $\Delta L=1$ scattering terms for RH neutrinos.
    \begin{itemize}
        \item $[N\leftrightarrow l\Phi_2]$ , $[N\leftrightarrow\Bar{l}\Phi_2^*]$
        \item $[ll\leftrightarrow\Phi_2^*\Phi_2^*]  ,  [l\Phi_2\leftrightarrow\Bar{l}\Phi_2^*] , [\Bar{l}\Bar{l}\leftrightarrow\Phi_2\Phi_2]$
    \end{itemize}
Now we can solve the above DE (\ref{3}) and (\ref{4}) as first-order linear in-homogeneous equations and find the solutions for $Y_{N}$ and $Y_{B-L}$ \cite{buchmuller2005pedastrian} as,\\
\begin{equation}
     Y_N(x)=\int dx^{'} DY^{eq}_Ne^{-\int dx^{''}D}+Y_N^{i}e^{-\int dx^{'}D}
\end{equation}

\begin{subequations}
    \begin{equation}
        Y_{B-L}(x)=Y_{B-L}^{i}e^{\int dx^{'}W} + \frac{3}{4}\epsilon_N\kappa
    \end{equation}
Where the efficiency factor $\kappa$ is given as, $0\leq\kappa\leq1$  
    \begin{equation}
        \kappa(x)=\frac{4}{3}\int dz^{'}D(Y_{N}-Y_{N}^{eq})e^{-\int dx^{''}W}
    \end{equation}
\end{subequations}
The interesting part about (\ref{4}) is the contribution of washout terms \cite{hugle2018low}, washout refers to the sweeping of the net lepton asymmetry due to decays and scatterings that might be generated, it should be noted that the $\Delta L=1$ scattering terms\footnote{For a much more detailed view refer \cite{buchmuller2005pedastrian} } is not been elaborated here, but will be considered for final calculations. Therefore the contribution to $W^{Total}$ comes from the inverse decays \cite{buchmuller2005pedastrian,luty1992baryogenesis,plumacher1997baryogenesis} and $\Delta L=2$ scattering terms denoted as,
\begin{equation}
    W^{Total}\simeq W_{ID}+W_{\Delta L=2}
\end{equation}
$$H(T)=1.66g^{\frac{1}{2}}_*T^2/m_{pl}=H(x=1)\frac{1}{x^2}$$
   $$ W_n=\frac{\Gamma_n}{Hx} $$
   $$\Gamma_{ID}=\Gamma_{D}\frac{Y^{eq}_{N_2}}{Y^{eq}_{l}}$$
   $$ W_{ID}=\frac{1}{4}Kx^3K_1(x)=\frac{1}{2}D(x)\frac{Y^{eq}_{N_2}}{Y^{eq}_{l}}$$
   $$W_{\Delta L=2}=\frac{36\sqrt{5}M_{Pl}}{\pi^{\frac{1}{2}}g_l\sqrt{g_{*}}v^4}\frac{1}{z_l^2}\frac{1}{\lambda_5^2}M_2\Tilde{m}^2$$
Where $g_l$ is the internal degree of freedom, $\Tilde{m}$ is the effective neutrino mass parameter \cite{borah2019tev} and $H$ is the Hubble parameter.
The $CP$ asymmetry comes from the loop level diagrams, at the tree level the $\Gamma_{Nl}$ and $\Gamma_{N\Bar{l}}$ are equal so the $CP$ asymmetry parameter is zero which means that there is no net lepton production. The two loop-level contributions are from the vertex correction and self-energy. So, according to Sakharov condition, we quantify the $C$ and $CP$ violation through a parameter $\epsilon$, and the sphalerons make the $B$ number violation and last condition of departure from thermal equilibrium at certain $T\sim M_i$ help the generation of asymmetry. The $CP$ asymmetry parameter\footnote{A much elaborated method for obtaining the $\epsilon$ is given in \cite{buchmuller1998cp}} is written as \cite{covi1996cp},\\
\begin{equation}
\epsilon_N=\frac{\Gamma_{N\xrightarrow{}l\Phi_2}-\Gamma_{N\xrightarrow{}\Bar{l}\Phi_2^*}}{\Gamma_{N\xrightarrow{}l\Phi_2}+\Gamma_{N\xrightarrow{}\Bar{l}\Phi_2^*}}\end{equation}\\
The tree level diagram doesn't contribute to the parameter as,
$$\Gamma_{N\xrightarrow{}l\Phi_2}=\Gamma_{N\xrightarrow{}\Bar{l}\Phi_2^*}=\frac{(Y^{\dagger}Y)_{ii}}{16\pi}$$
The Feynman rules for the interactions of right-handed neutrinos are given in \cite{gluza1992feynman,denner1992feynman} and Therefore, the only possible contribution from higher-order diagrams are,
  $$ 
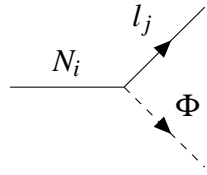
\begin{tikzpicture}
    \begin{feynman}
                 \vertex(a);
    \vertex[ left=of a](b){};
    \vertex[above right =of a](c);
        \vertex[ below right=of a](d);
    \diagram{
    (b) -- [edge label=$N_i$](a);
    (a) -- [fermion, edge label=$l_j$](c);
    (a) -- [dashed,fermion, edge label=$\Phi$](d);
     };
    \end{feynman}
\end{tikzpicture}$$
$$(\epsilon_{tree-level})_N=0
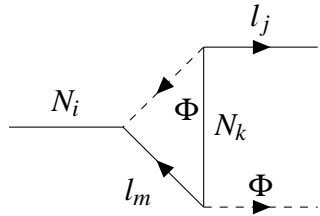
$$
$$
\begin{tikzpicture}
    \begin{feynman}
                 \vertex(a);
    \vertex[ left=of a](b){};
    \vertex[above right =of a](c);
        \vertex[ below right=of a](d);
  \vertex[  right=of c](e){};
  \vertex[ right= of d](f){};
    \diagram{
    (b) -- [edge label=$N_i$](a);
    (c) -- [dashed,fermion,edge label=$\Phi$](a);
    (d) -- [fermion, edge label= $l_m$](a);
    (d) -- [edge label'=$N_k$](c);
    (c) -- [fermion,edge label=$l_j$](e);
    (d) -- [dashed,fermion, edge label=$\Phi$](f);

     };
    \end{feynman}
\end{tikzpicture}$$
\begin{equation}
    (\epsilon_{vertex})_N=\frac{1}{8\pi}\sum\frac{\text{Im}\left[\left(Y^{\dagger}Y\right)^2_{ki}\right]}{\left(Y^{\dagger}Y\right)_{ii}}\times\frac{M_k}{M_i}\left(1-(1+\frac{M_k^2}{M_i^2})\ln[(M_i^2+M_k^2)/(M_k^2)]\right)
\end{equation}\\
The $C$ and $CP$ violations is arising in the form of complex terms in this parameter \cite{botella,kolb2018early} and with the help of unitarity relation \cite{kolb1980baryon} one can obtain the desired result.
$$
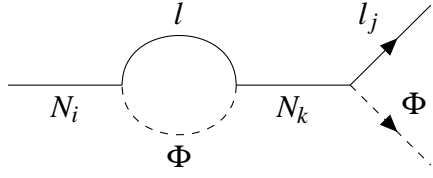
\begin{tikzpicture}
    \begin{feynman}
                 \vertex(a);
    \vertex[ left=of a](b){};
    \vertex[ right =of a](c);
        \vertex[  right=of c](d);
  \vertex[ above right=of d](e);
  \vertex[ below right= of d](f);
    \diagram{
    (b) -- [edge label'=$N_i$](a);
    (d) -- [fermion, edge label=$l_j$](e);
    (c) -- [ half right,edge label'=$l$](a);
    (c) -- [dashed, half left, edge label=$\Phi$](a);
    (c) -- [edge label'=$N_k$](d);
    (d) -- [dashed, fermion, edge label=$\Phi$](f);
     };
    \end{feynman}
   \end{tikzpicture}$$ 
\begin{subequations}
\begin{equation}
(\epsilon_{self-energy})_N=-\frac{1}{8\pi}\sum_{k\neq i}\frac{M_i}{M_k^2-M_i^2}\frac{\text{Im}\left[({M_k(Y^{\dagger}Y)_{ki}+M_i(Y^{\dagger}Y)_{ik})}Y^*_{jk}Y_{ji}\right]}{(Y^\dagger Y)_{ii}}
\end{equation}
\begin{equation}
\simeq-\frac{1}{8\pi}\sum_{k\neq i}\frac{\text{Im}\left[\left(Y^{\dagger}Y\right)^2_{ki}\right]}{\left(Y^{\dagger}Y\right)_{ii}}\frac{M_iM_k}{M_k^2-M_i^2}
\end{equation}
\end{subequations}

\begin{equation}
    \epsilon_N=(\epsilon_{vertex})_N+(\epsilon_{self-energy})_N\simeq-\frac{3}{16\pi}\sum_{k=1,3}\frac{M_i}{M_k}\frac{\text{Im}\left[\left(Y^{\dagger}Y\right)^2_{k2}\right]}{\left(Y^{\dagger}Y\right)_{22}}
\end{equation}

$$\Gamma_i=\frac{M_i}{8\pi}\left(Y^{\dagger}Y\right)_{ii}\left(1-\frac{(m_\Phi)^2}{(M_i)^2}\right)$$
$$Y^+_vY_v=UD_{\sqrt{\kappa}}O^+D_{M}RD_{\sqrt{\kappa}}U^+$$
Now we look to relate the baryon number with the excess of lepton number, $B$ and $B-L$ . Particle asymmetries can be related by chemical potentials, the relation between excess of the particle over its antiparticle with chemical potential, with $g$ as the number of internal degrees of freedom \cite{harvey1990cosmological},
$$n_+-n_-=\frac{gT^3}{3}\left[\frac{\mu}{T}\right] (\text{Bosons}) $$
$$n_+-n_-=\frac{gT^3}{6}\left[\frac{\mu}{T}\right] (\text{Fermions})$$
Now we can relate the quantum number with a number density $n$ as $B=\frac{n_b-n_{\Bar{b}}}{s}$ where $s$ is the entropy density and is proportional to $R^{-3}$. So we can now write the equations with $3+N$, where $N$ (total number of generations) and $m$ (number of complex Higgs doublet) chemical potentials for the above-mentioned processes also keeping the charge of the universe neutral, in total, we get five independent equations. After solving for $B$ and $L$ we get,\\
\begin{equation}
\label{7}
    B=\frac{8N+4m}{22N+13m}(B-L)
\end{equation}
\begin{equation}
\label{8}
    L=-\frac{14N+9m}{22N+13m}(B-L)
\end{equation}
\begin{equation*}
    (B+L)=-\frac{6N+5m}{22N+13m}(B-L)
\end{equation*}
 The baryon asymmetry in the universe is expressed as the baryon-to-photon ratio $\eta$. The entropy degrees of freedom are defined as $g_s$ and the ratio between the initial and final degrees of freedom is called dilution factor $f$ \cite{buchmuller2002cosmic}, for our model we have $a_{sph}\simeq 0.34$ and Baryon-to-photon ratio in terms of $B-L$ as,
\begin{equation}
\label{9}
    \eta=\frac{n_B(t_0)}{n_{\gamma}(t_0)} = \frac{1}{f}N^0_B = \frac{1}{f}\frac{3}{4}\epsilon_D \kappa
\end{equation}
\begin{equation}\label{90}\eta=\frac{3}{4}\frac{g_s^0}{g_s^*}a_{sph}\epsilon_N\kappa \simeq 0.013N^0_{B-L}\end{equation}
So, when the sphalerons are in equilibrium there is a conversion of lepton to baryon number ($n_L=-a_{sph}n_{B-L}$), the factor of $a_{sph}$ is sphalerons conversion factor which is responsible for conversion of different quantum number. When at very high temperatures there is no such equilibrium, we have $n_{B-L}=-n_L$ \cite{buchmuller1999matter}, and from (\ref{90}) we get the observed baryon-to-photon ratio of the universe.
\subsection{\label{concl}Conclusion}
In this literature, the aim has been to give a simpler idea about some well-observed phenomena that were not explained in the Standard model of physics. By no means should it be understood that the standard model is wrong it is just incomplete, to explain these phenomena we have models beyond the standard model, one such model is the scotogenic model. Working within the scotogenic framework as a common origin, I try to explain dark matter, neutrino mass and matter-antimatter asymmetry. For CDM we choose the neutral component of the second Higgs doublet which due to an additional $Z_2$ symmetry is stable, this DM candidate gives the observed value for the relic abundance of Dark matter. The next aspect is the generation of neutrino mass at the loop level from the self-energy diagram, for which we integrate out our right-handed neutrinos. The advantage of this mechanism over the seesaw is the aspect of verification in current experiments, the small mass can be obtained for neutrinos with the values for Yukawa coupling, and $\lambda_5$ we do not dwell deeper in the constraining and putting bounds over masses and constants. Next, the aspect of baryon asymmetry that is observed as the Baryon-to-photon ratio $\eta$, is explained through thermal leptogenesis. There are other methods such as baryogenesis at the GUT scale that explain the asymmetry, but the idea for baryon-asymmetry from lepton-asymmetry (leptogenesis) does not involve any supersymmetry and GUT, which makes it much simpler. Now to explain baryogenesis we start with fulfilling the Sakharov conditions and you will see that to quantify the $C$ and $CP$ violation we use $\eta$ parameter, thereafter the procedure is simpler as initially, we write the Boltzmann equation for the evolution of particular species, secondly, the parameters related to the B.E such as $\eta$, the number density, and various scattering cross-sections that are related to net lepton asymmetry are calculated. Finally, we consider the transitions, that are made possible due to fluctuations at the thermal bath which converts the excess lepton number to baryon number, to quantify this we use chemical potentials, and at last, we obtain a relation between $n_{L}$ and $n_{B-L}$ with a factor $a_{sph}\simeq0.34$ which governs the conversion.  This model is a common origin for explaining DM, neutrino masses and matter-antimatter asymmetry, but there are certain aspects that this model doesn't aim to explain such as the strong $CP$ problem but this doesn't make this framework any less significant with all testable advantages.
\subsection*{Acknowledgement}
I wish to thank $\text{Dr.}$ Abhass Kumar\footnote{Department of Physics and Astrophysics, University of Delhi-110021} with whom a series of discussions on various models beyond the standard model helped me in drafting this report.
\bibliographystyle{ieeetr}
\bibliography{z}

\end{document}